%% file: template.tex
\documentclass{article}

\usepackage{arxiv}

\usepackage{hyperref}       %
\usepackage{amsmath}
\usepackage{multirow}

\usepackage{amsmath,amsfonts}
\usepackage{algorithm}
\usepackage{array}
\usepackage[caption=false,font=normalsize,labelfont=sf,textfont=sf]{subfig}
\usepackage{textcomp}
\usepackage{stfloats}
\usepackage{url}
\usepackage{verbatim}
\usepackage{graphicx}
\usepackage{cite}
\usepackage{hyperref}       %
\usepackage{MnSymbol}
\usepackage{multirow}

\usepackage{algorithm}
\usepackage{algpseudocode}

\usepackage{pgfplots}
\usepgfplotslibrary{colorbrewer}
\pgfplotsset{compat = 1.15, cycle list/Set1-9} 
\usetikzlibrary{pgfplots.statistics, pgfplots.colorbrewer} 
\usepackage{pgfplotstable}

\usepackage{algorithm}
\usepackage{algpseudocode}

\usepackage{pgfplots}
\usepgfplotslibrary{colorbrewer}
\pgfplotsset{compat = 1.15, cycle list/Set1-9} 
\usetikzlibrary{pgfplots.statistics, pgfplots.colorbrewer} 
\usepackage{pgfplotstable}

\title{Breaking XOR Arbiter PUFs with Chosen Challenge Attack}

\usepackage{authblk}

\author[1,2]{%
{Niloufar Sayadi}%
}
\author[3]{%
{Phuong Ha Nguyen}%
}\author[1,2,4]{%
{Marten van Dijk}%
}
\author[1]{%
{Chenglu Jin}%
}
\affil[1]{CWI Amsterdam, The Netherlands}
\affil[2]{Vrije Universiteit Amsterdam, The Netherlands}
\affil[3]{eBay, CA, United States}
\affil[4]{University of Connecticut, CT, United States}

\renewcommand{\shorttitle}{Breaking XOR Arbiter PUFs with Chosen Challenge Attack}

\hypersetup{
pdftitle={Breaking XOR Arbiter PUFs with Chosen Challenge Attack},
pdfauthor={Niloufar~Sayadi, Phuong~Ha~Nguyen, Marten~van~Dijk and Chenglu~Jin},
pdfkeywords={XOR Arbiter PUFs, PUF Modeling Attacks, Chosen Challenge Attacks, Reliability-based Attacks},
}

\begin{document}
\maketitle

\input{sections/abstract}

\keywords{XOR Arbiter PUFs \and PUF Modeling Attacks \and Chosen Challenge Attacks \and Reliability-based Attacks}

\input{sections/intro}
\input{sections/background}

\input{sections/attack}
\input{sections/experiments}
\input{sections/countermeasure}

\input{sections/relatedwork}

\input{sections/conclusion}

\section*{Acknowledgements}
We would like to thank the valuable comments from the anonymous reviewers. Marten van Dijk and Chenglu Jin are (partially) supported by project CiCS of the research programme Gravitation which is (partly) financed by the Dutch Research Council (NWO) under the grant 024.006.037.

\bibliographystyle{plain}

\appendix
\input{sections/Appendix}

\end{document}

%% file: sections/abstract.tex
\begin{abstract}

The XOR Arbiter PUF was introduced as a strong PUF in 2007 and was broken in 2015 by a Machine Learning (ML) attack, which allows the underlying Arbiter PUFs to be modeled individually by exploiting reliability information of the measured responses. To mitigate the reliability-based attacks, state-of-the-art understanding shows that the reliability of individual Arbiter PUFs and the overall XOR Arbiter PUF can be boosted to an arbitrarily high level, thus rendering all known reliability-based ML attacks infeasible; alternatively, an access control interface around the XOR Arbiter PUF can prevent the same challenge-response pairs from being accessed repeatedly, thus eliminating the leakage of reliability information.

We show that, \textbf{for the first time, a perfectly reliable XOR Arbiter PUF can be successfully attacked in a divide-and-conquer manner}, meaning each underlying Arbiter PUF in an XOR Arbiter PUF can be attacked individually. This allows us to attack large XOR Arbiter PUFs efficiently, even without reliability information or any side-channel information. Our key insight is that, instead of reliability information, the responses of highly correlated challenges also reveal how close the responses are to the response decision boundary. This leads to a \textit{chosen challenge attack} on XOR Arbiter PUFs by carefully choosing correlated challenges to measure and aggregate the collected information. We validate our attack by using PUF simulation, as well as an XOR Arbiter PUF implemented on FPGA. We also demonstrate that our chosen challenge methodology is compatible with the state-of-the-art combined gradient-based multi-objective optimization attack. Finally, we discuss an effective countermeasure that can prevent our attack but with a relatively large area overhead compared to the PUF itself.

\end{abstract}

%% file: sections/intro.tex
\section{Introduction}
The unique process manufacturing variation on every chip can be leveraged to create a Physical Unclonable Function (PUF) as a fingerprint for the device~\cite{gassend2002silicon}. A vector of bits is applied to the PUF as the PUF challenge, and a unique response is generated by the PUF as output. PUFs are useful for device authentication and secret key management~\cite{lim2005extracting,suh2007physical,gurevin2023secure}.

One classical design of silicon PUFs is Arbiter PUF (APUF), which measures the delay differences of two competing paths determined by a challenge and produces a one-bit response depending on which path is faster~\cite{gassend2002silicon}. Ideally, the challenge-response pairs (CRPs) of an APUF are only determined by process variation and thus unpredictable to an attacker. However, the behavior of any APUF can be easily modeled by machine learning (ML) attacks after collecting many CRPs of the APUF~\cite{ruhrmair2010modeling}. A precise mathematical model of a PUF allows the attacker to violate the security properties of the PUF and thus successfully impersonate a legitimate user in authentication or retrieve secret keys managed by the PUF.

XOR Arbiter PUF (XOR APUF) was introduced as a security enhancement of the APUF design. It XORs multiple parallel APUF response bits together to form a response bit of the XOR APUF. In addition to being a strong component of many state-of-the-art strong PUFs, e.g. iPUF~\cite{nguyen2018interpose}, XOR APUF is one of the few PUFs with practical applications such as PUF-based RFID tags~\cite{becker2015gap}. XOR APUF has been considered a promising candidate for secure strong PUF design since their introduction in 2007~\cite{suh2007physical} until Becker presented a reliability-based ML modeling attack that can model the individual underlying APUFs in 2015~\cite{becker2015gap}. The reliability-based attack relies on the fact that the reliability of a CRP under the measurement noise reveals whether the response is close to the response decision boundary where the value of the response bit will be flipped. This extra information allows attackers to implement a divide-and-conquer strategy to attack individual APUFs in an XOR APUF rather than the whole XOR APUF~\cite{becker2015gap}.  
    
One limitation of the reliability-based attack is that it has to access the reliability information of individual CRPs by repeated measurement. Following this line of thinking, countermeasures have been proposed to thwart or prevent the leakage of reliability information by designing super reliable XOR APUFs~\cite{wisiol2019attackers} or blocking repeated measurements using an access control interface~\cite{jin2020erasable}. 

\textbf{In this paper, we introduce a novel \textit{divide-and-conquer} strategy to model the XOR APUF without the reliability information of CRPs or any side-channel information of the PUF}. Therefore, the XOR APUF design is completely broken in both reliable and unreliable cases. Note that the proposed attack requires the attackers to choose the challenges applied to the PUF; however, it does not mean that we assume a stronger adversarial model than the reliability-based attack because the reliability-based attack is also one kind of chosen challenge attack, as it chooses to measure the same CRPs repeatedly.

In our attack, we choose a few challenges around a randomly selected challenge with only a one-bit difference in their $\Psi$ vectors (a $\Psi$ vector is a transformed challenge based on Eq.~\ref{phi_eq}), and then we check how many surrounding challenges will lead to a flipped response bit from the response of the centroid challenge. The flipping probability reveals how close the centroid challenge is to the response decision boundary, where the response bit flips. Relying on this fact, we propose a chosen challenge attack that can model individual APUFs in an XOR APUF and thus break the security of XOR APUF completely, even in the case when state-of-the-art reliability-based attacks fail. %

Our experimental results in Sec.~\ref{exp_reli} show that the proposed attack can model individual APUFs in a perfectly reliable XOR APUF with a high prediction accuracy, while the reliability-based attacks fail to work~\cite{becker2015gap}. As shown in our experiments in Sec.~\ref{exp_unreli}, our attack can also model the XOR APUF when the PUF is realistically unreliable. %
Finally, we also validate our attack by attacking XOR APUFs implemented on a real FPGA. %

\subsection{Contributions}
We make the following contributions in this paper: 
\begin{itemize}
    \item We introduce a novel chosen challenge attack on unreliable and, more importantly, perfectly reliable XOR-Arbiter PUFs.
    \item We evaluate the effectiveness of the proposed attack using a PUF simulation under various conditions, including various noise levels, sampling rates, and the number of flips.

    \item We evaluate the compatibility of the proposed chosen-challenge methodology with the latest combined Logistic Regression (LR) and reliability-based attack proposed in~\cite{tobisch2021combining} and provide a combined LR non-flipping-probability-based attack, which is applicable for both reliable and unreliable PUF.
    
    \item We demonstrate the practicality of our attack by attacking a real XOR APUF implemented on an FPGA.
    \item We discuss an effective countermeasure against our attack.  %
    \item As a proof of concept, the source code of our Chosen Challenge attack is publicly available on GitHub.\footnote{\url{https://github.com/niloufarsyd/Chosen_Challenge_Attack}}

\end{itemize}

\subsection{Organization}
We present the necessary background of APUF, XOR APUF, and the reliability-based attacks in Sec.~\ref{background}. The proposed chosen challenge attack is discussed in Sec.~\ref{attack}. Sec.~\ref{exp} presents the experimental results and a fair comparison with the state-of-the-art reliability-based attacks. An effective countermeasure is discussed in Sec.~\ref{Countermeasures}, followed by some other related work in Sec. \ref{relatedwork}. The paper concludes in Sec.~\ref{conclusion}.

%% file: sections/background.tex
\section{Background}\label{background}
\subsection{Arbiter PUFs and XOR Arbiter PUFs}
An Arbiter PUF (APUF) is a strong PUF that consists of $n$ consecutive delay stages that lead to an arbiter. The stages are identical 2-to-1 multiplexers that lead the top and bottom signals based on the challenge bits that are applied to their $select$ inputs. In the last stage, the top and bottom signals have an accumulated delay difference due to the differences in the delay of every stage introduced by process variations. Finally, the arbiter outputs a `1' or `0' bit depending on which signal arrives at the arbiter earlier. %

The behavior of an APUF can be captured by a {\em Linear Additive Delay Model}~\cite{lim2005extracting} 
\begin{equation} \label{delta_eq}
\begin{split}
\Delta =w[0]\Psi[0]+...+w[n]\Psi[n]= \langle W , \; \Psi \rangle 
\end{split},
\end{equation}
where $W$ is the weight vector,  $\Psi$ is the parity (or feature) vector and $n$ is the number of challenge bits or the number of delay stages. The weight vector $W$ defines the character of the APUF, and it is determined only by process variations. For $\Delta\geq 0$, the response $r=1$, otherwise $r=0$. The parity vector $\Psi$ is solely based on the challenge vector $c$ in the following way: 
\begin{equation} \label{phi_eq}
\begin{split}
\Psi[n]=1 \hspace{4pt} and \hspace{4pt} \Psi[i]=\prod_{j=i}^{n-1}(1-2c[j]) , i=0,...,n-1.
\end{split}
\end{equation}

A $k$-XOR APUF is constructed by $k$ APUFs that are fed by the same challenge, and their response bits are XORed together. Unlike APUFs, XOR APUFs have a non-linear model due to the added XOR. Thus, it was generally believed that XOR APUFs are much harder to model than APUFs, and the difficulty grew exponentially in $k$ before the reliability-based attack was known~\cite{ruhrmair2010modeling}. %

\subsection{Reliability-Based Attacks on XOR APUFs}
Unlike the classical machine learning modeling attacks on PUFs that use response bits directly for training, in a reliability-based modeling attack, the reliability information derived from repeated measurement of CRPs is used for modeling. The response to each challenge is measured multiple times to compute the reliability of the CRP. This reliability gives information about the delay difference $\Delta$ of an APUF component in an XOR APUF under the evaluated challenge. This is because when $|\Delta|$ is smaller than a threshold value $\epsilon$, the CRP is unreliable, and the response can easily be flipped between $0$ and $1$; otherwise, the CRP is reliable, and the response is consistent. Also, in an XOR APUF, if one of the underlying APUFs is unreliable, then the overall XOR APUF is also unreliable. This is why the reliability information reveals the delay information of the individual APUF and allows a divide-and-conquer strategy to model individual APUFs in an XOR APUF. %
In~\cite{becker2015gap}, Becker used a popular evolution strategy-based optimization algorithm, Covariance Matrix Adaptation Evolution Strategy (CMA-ES), to learn the weight vector of the APUF along with the threshold value $\epsilon$. For this purpose, a set of challenges $C_i$ is randomly generated, and the reliability $R_i$ is measured for every challenge using Eq.~\ref{Reli1}, where $M$ is the number of measurements per CRP. %
\begin{equation}\label{Reli1}
    R_i=|\frac{M}{2}-\sum_{i=1}^{M} {r_i}|
\end{equation}

Then, Becker used CMA-ES to optimize the PUF model by maximizing the Pearson correlation coefficient of the measured reliability and the predicted reliability based on the PUF models in the current iteration of the CMA-ES algorithm. The algorithm will converge to the model of an underlying APUF in the attacked XOR APUF. 
By executing the CMA-ES algorithm multiple times, all APUF models in the XOR APUF will be modeled eventually. %

An enhanced reliability-based attack on XOR APUF was presented by Nguyen \textit{et al.}, which utilizes a more accurate reliability model to improve the prediction accuracy or attack efficiency~\cite{nguyen2018interpose}. Tobisch \textit{et al.} demonstrated that it is also possible to use a gradient-based optimization to launch reliability-based attacks and to combine the objective of the reliability-based attack and that of the classical CRP-based attack~\cite{tobisch2021combining}. %

%% file: sections/attack.tex
\section{Proposed Attack on XOR APUFs}\label{attack}
\newcommand{\ignore}[1]{}
\subsection{Motivation}\label{attack_motiv}
To the best of our knowledge, the reliability-based attack is still the most efficient attack on XOR APUFs because the attackers can exploit more fine-grained information than just the CRPs used in classical ML attacks. %
However, if no reliability information is available to the attacker, then none of the reliability-based attacks~\cite{becker2015gap,nguyen2018interpose,tobisch2021combining} would work, and the attackers will have to use the best-known classical attacks~\cite{mishra2024calypso,ruhrmair2013puf,santikellur2019deep,santikellur2020computationally,ruhrmair2010modeling,wisiol2022neural}. 

As shown in~\cite{wisiol2019attackers} \textcolor{black}{and~\cite{anandakumar2022implementation}}, by implementing a majority voting at the end of every APUF in an XOR APUF, one can boost the reliability of the APUFs and the overall XOR APUF arbitrarily. Although this defense may incur a performance overhead, it is very effective in mitigating the reliability-based attack because the attacker cannot find a noisy CRP to derive its reliability. \textcolor{black}{Furthermore, the challenge obfuscation proposed in~\cite{anandakumar2022implementation} is based on DES permutation, which is reversible and known to adversaries.} %
Another potential countermeasure that could paralyze the reliability-based attacks is the (logically) erasable PUF interface~\cite{jin2020erasable,jin2022programmable}. %
Using such an interface, the access to each individual CRP can be controlled, and then, it is impossible to remeasure the response of an erased challenge anymore, e.g., the interface can automatically erase the measured challenge after one or a certain number of measurements. Thus, it prevents the attacker from deriving the reliability information even if the underlying PUF is still noisy. %
Our proposed attack can defeat both of the countermeasures since we do not rely on the reliability information anymore. %

\subsection{Adversarial Model}
We assume the same adversarial model as the reliability-based attack~\cite{becker2015gap}. The attackers can apply any challenges they want to the PUF and only get the responses of the queried challenges back, i.e., no other side channel information, like power or timing information~\cite{ruhrmair2014efficient}. The only difference is that we \textit{choose} to measure correlated CRPs, but the reliability-based attacks \textit{choose} to measure the same CRPs repeatedly.

\subsection{Chosen Challenge Attack}\label{attack_main}

\noindent\textbf{Key Idea.} In this strategy, %
we focus on manipulating $\Psi$ and investigate how it affects the delay model (Eq.~\ref{delta_eq}).
Suppose that there is a flip in the $\Psi$ vector of the delay model; then, this flip will lead to a different coefficient for one of the weight vector elements. Note that a challenge $c$ consists of only $0$ or $1$, but a $\Psi$ vector only contains $1$ or $-1$, according to Eq.~\ref{phi_eq}. Thus, after one flip in $\Psi$ (from $1$ to $-1$ or from $-1$ to $1$), the delay difference $\Delta$ would be changed from $\Delta_1$ to $\Delta_2$, as shown in Eq.~\ref{delta1_eq}. Whether this change leads to a flipped response bit reveals whether $\Delta$ flips its sign, i.e., whether $|\Delta| < 2|w[1]|$, in the case of Eq.~\ref{delta1_eq}. This is the key enabler of our attack on APUFs. 
\begin{equation} \label{delta1_eq}
\begin{split}
\begin{cases}
\Delta_1 =w[0]\textcolor{red}-w[1]-w[2]+...+w[n]\\
\Delta_2 =w[0]\textcolor{red}+w[1]-w[2]+...+w[n]
\end{cases}
\end{split}
\end{equation}

This relation also allows us to attack individual APUFs in an XOR APUF. \ignore{Statistically speaking}According to the analysis provided in Appendix~\ref{AppendixA}, if there is a flip in $\Psi$, it is unlikely that the response of an APUF will be flipped. But if the response of one APUF is flipped, this flip will be propagated to the output of the XOR and be observable by the attacker, assuming this flip in $\Psi$ does not flip the responses of the other APUFs. Of course, multiple APUF responses can be flipped occasionally, but we will use the Pearson correlation coefficient as a robust indicator to deal with the ``noise'' introduced by multiple APUF response flips. %

\begin{algorithm}[t]
\caption{Chosen Challenge Attack}\label{alg:Flip}

\begin{algorithmic}[1]

\Procedure{Chosen Challenge Attack}{}
\For{$t \leftarrow 1$ to $N$} \Comment{Training Data Collection}
    \State randomly select a challenge $c$
    \State transform $c$ into $\Psi$ based on Eq.~\ref{phi_eq}
    \State $r \leftarrow PUF(c)$
    \For{$i \leftarrow 1$ to $m$}
        \State randomly generate $\Psi_{i}$ s.t. $d_H(\Psi,\Psi_{i})=1$
        \State transform $\Psi_i$ into $c_{i}$ by inverse Eq.~\ref{phi_eq}
        \State $r_{i} \gets PUF(c_{i})$
    \EndFor
    \State $F_t=1-\frac{\sum_{i=o}^{m-1}|r_{i}-r|}{m}$
    \State $\mathcal{Z} = \{(\Psi_i, F_i)\}_{i=1}^{N}$
\EndFor

\State generate K random models:
\{$W_1$,...,$W_j$,...,$W_K$\}
\For{$j \leftarrow 1$ to $K$} \Comment{Optimization}
        \For{$h \leftarrow 1$ to $N$}
            \State $F'_{j,h}=|\Delta| =|\langle W_j , \; \Psi_{h} \rangle|$ 
        \EndFor
    \State $F'_j = \{F'_{j,h}\}^N_{h=1}$
    \State $\rho_j = Correlation(F,F_j')$ %
\EndFor
    \State CMA-ES uses $L$ out of the $K$ models of $W$ with the highest $\rho$ to generates K new models.
\State repeat lines (15-22) for $T$ iteration to maximize $\rho$ and output the optimized model $W$.
\EndProcedure
\end{algorithmic}
\end{algorithm}

\vspace{2mm}
\noindent\textbf{Detailed Steps.} Algorithm \ref{alg:Flip} shows the pseudo-code of the proposed chosen challenge attack, and it models the PUF using the flipping probability information. To compute the flipping probability, we first choose a challenge $c$ randomly in the challenge space and then convert it to the corresponding $\Psi$ according to Eq.~\ref{phi_eq}. Then we randomly sample $m$ $\Psi_{i}$ vectors that have a Hamming distance of 1 from $\Psi$:
\begin{equation} \label{HD_eq}
\begin{split}
d_H(\Psi,\Psi_{i})=1.
\end{split}
\end{equation}

The Hamming distance of $1$ means only one bit is different between $\Psi$ and $\Psi_i$. Practically speaking, if one bit $\Psi[i]$ is flipped, then two consecutive bits ($c[i-1]$ and $c[i]$) should be flipped in the corresponding challenge $c$, except for $i = 0$, then only $c[0]$ needs to be flipped. 

\begin{figure}[t]
    \centering
    \includegraphics[width=.5\columnwidth]{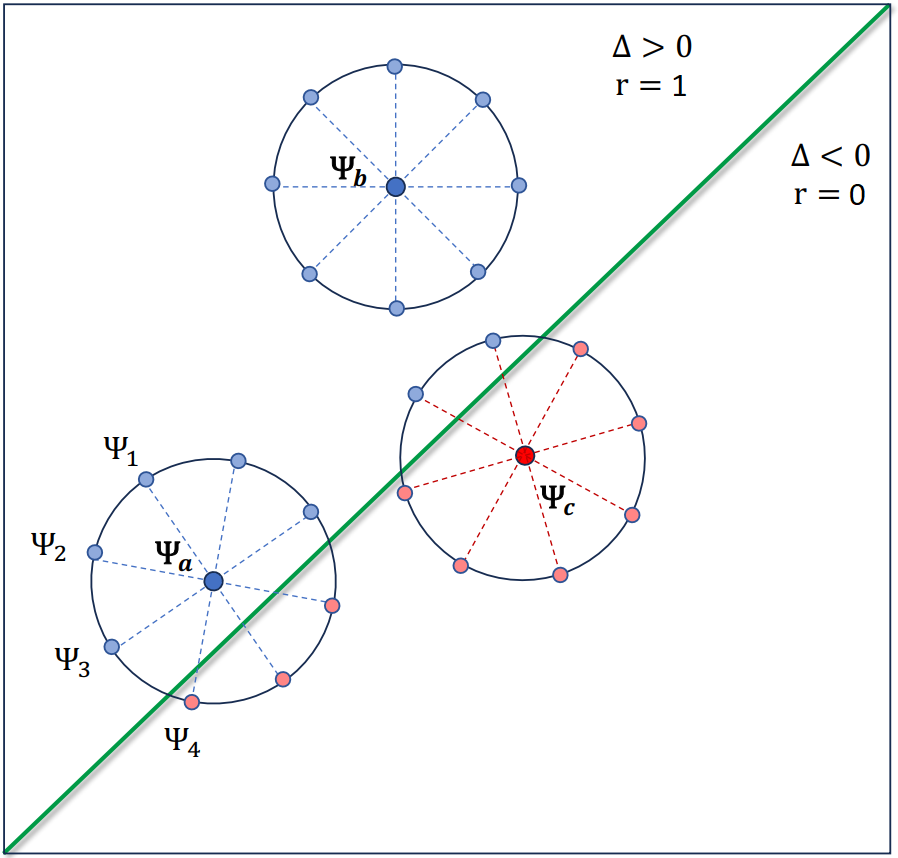}
    \caption{Geometric illustration of the chosen $\Psi$ vectors and their relation with the linear response decision boundary in the $\Psi$ space. Note that this diagram is for illustration purposes only, as the $\Psi$ space is never a 2-D space. The decision boundary is always linear due to the linear delay model in Eq.~\ref{delta_eq}.}
    \label{fig:AdaptiveChallenge}
\end{figure}

Fig.~\ref{fig:AdaptiveChallenge} illustrates the main idea of our attack. For any randomly chosen $\Psi$ (e.g., $\Psi_{a}$, $\Psi_{b}$, or $\Psi_{c}$), the $m$ sampled points with $d_H=1$ are on the circular radius to the centroid of $\Psi$. If the $\Psi$ is close to the response decision boundary, it is more likely that the responses of the sampled neighboring $\Psi_i$ vectors are different from the response of $\Psi$. %

Note that in Algorithm~\ref{alg:Flip} and in our source code, the non-flipping probability value is used. The non-flipping probability is calculated as in Eq.~\ref{flip_eq}.
\begin{equation} \label{flip_eq}
\begin{split}
F=1-\frac{\sum_{i=1}^{m}|r_{i}-r|}{m}
\end{split}
\end{equation}
The responses $r_{i}$ and $r$ in the Eq.~\ref{flip_eq} are the PUF responses of the sampled neighboring $\Psi_i$ and the centroid $\Psi$, respectively, and $m$ is the number of sampled $\psi_i$ around a centroid $\Psi$. %

Thus, after sampling many random challenges and their surrounding challenges, we get a dataset of $(\Psi, F)$ pairs as Eq.~\ref{FPSet}.
\begin{equation}\label{FPSet}
\mathcal{Z}=\{(\Psi_1, F_1),...(\Psi_i, F_i),...,(\Psi_N, F_N)\},   \end{equation}
where $N$ is the size of the training dataset in our attack.

Then, the CMA-ES algorithm is used to find the optimal $W$ model for individual APUF in an XOR APUF. To start the procedure, $K$ random models are generated. For each model $W_j$, the $\Delta$ is calculated for each $\Psi$ vector included in $\mathcal{Z}$, and the predicted non-flipping probability $F^{'}$ is computed according to Eq.~\ref{flip_pred}. The expected linear relationship of $F$ and $F'$ will be discussed in Sec.~\ref{theoritical}.
\begin{equation}\label{flip_pred}
    F^{'}=|\Delta| = |\langle W , \; \Psi \rangle|
\end{equation}

The objective function of CMA-ES is the Pearson correlation coefficient $\rho_j$ between the measured non-flipping probability ($F_1$,...,$F_N$) and the predicted non-flipping probability ($F^{'}_1$,...,$F^{'}_N$). Inside the CMA-ES algorithm, an $L$ number of $w$ APUF models with the highest $\rho_j$ value are kept in each iteration to generate a new generation of $K$ models. After $T$ iterations, the CMA-ES outputs the model with the highest Pearson correlation coefficient $\rho$ as the best model. When attacking an XOR APUF, the CMA-ES algorithm would converge to one of the individual APUFs in the XOR APUF. We can repeat the whole process, and eventually, all the APUFs of an XOR APUF will be modeled after a sufficient number of CMA-ES runs; therefore, the whole model of XOR APUF is revealed. 

\vspace{1mm}
\noindent \textbf{Time Complexity.} The modeling time complexity of the proposed Chosen Challenge Attack is shown in Eq.~\ref{time_complexity}. The timing complexity is determined by the number of centroid challenges or training challenge sets ($N$), the number of challenges sampled around each centroid challenge ($m$), the number of CMA-ES iterations ($T$), the number of models populated in each CMA-ES generation ($K$), and the size of $\Psi$ ($n$). The term $Nm$ in the time complexity is due to the training data collection phase of the attack, and the term $TKNn$ is due to the modeling optimization phase, as we can see in Algorithm~\ref{alg:Flip}.
\begin{equation}\label{time_complexity}
    \mathcal{O}(Nm+TKNn)
\end{equation}

\subsection{A Unified Theoretical Foundation of Our Attack and the Reliability-based Attacks}\label{theoritical}

As shown in~~\cite{delvaux2013side,nguyen2018interpose}, the non-flipping probability $F$ (or reliability) of a CRP in an APUF, under some perturbations, has a relationship with the delay difference $\Delta$ of the CRP and the standard deviation  $\sigma$ of the perturbations as follows.
 \begin{equation}\label{F_linear1}
   \Delta/ \sigma=\sum_{i=0}^{n} {(w[i]/\sigma)\Psi[i]}=-\Phi^{-1}(F)
\end{equation}

In a reliability-based attack, the perturbations come from environmental noise, and thus $\sigma$ is the standard deviation of the environmental noise with a normal distribution $\mathcal{N}(0,\sigma^2)$. Similarly, in our proposed chosen challenge attack, the perturbations come from the flipped $\Psi$ bit, so $\sigma$ is the standard deviation of the perturbation ($2W[i]$) caused by one flip in $\Psi$. Note that each weight vector component $W[i]$ in an APUF is believed to follow a normal distribution $\mathcal{N}(0,\sigma^2/4)$ if we define $\sigma$ to be the standard deviation of the random perturbation in this attack. This normal distribution assumption of the PUF weights $W[i]$ is widely used in APUF simulations and is validated in real APUF implementations in the past research~\cite{becker2015gap,nguyen2018interpose,ruhrmair2010modeling,ruhrmair2013puf,mishra2024calypso}.

Eq.~\ref{F_linear1} can be further approximated as Eq.~\ref{F_linear2} if $F\in[0.1,0.9]$.
\begin{equation}\label{F_linear2}
    |\Delta/ \sigma| \approx F
\end{equation}

Due to the linear relationship between the measured non-flipping probability $F$ and the delay model $\Delta$, which is also the predicted non-flipping probability $F'$, we can select the best-fitting models among random models and then optimize them in every iteration of the CMA-ES algorithm to generate an accurate model of an underlying APUF. %

\subsection{Comparison with Reliability-based Attacks}
The proposed attack shares many similarities with the reliability-based attack in~\cite{nguyen2018interpose}, which is an enhanced version of the attack in~\cite{becker2015gap}. Both our attack and the reliability-based attacks need to choose the CRPs to be measured, and they all exploit the fact that the non-flipping probability (or the reliability) has a linear relationship with the delay difference $\Delta$. Most importantly, both our attack and the reliability-based attack can learn individual APUFs in an XOR APUF in a divide-and-conquer manner. %

Actually, the proposed chosen challenge attack can be viewed as a generalization of the reliability-based attack. As we will show in Sec.~\ref{exp_2flip}, our chosen attack does not have to restrict the hamming distance to be 1; the known reliability-based attacks are a special case of our chosen challenge attack when the hamming distance is 0, and the perturbation is caused by small environmental noise instead of flips in $\Psi$. This is why our attack methodology is more widely applicable, including attacking perfectly reliable XOR APUFs. 

%% file: sections/experiments.tex
\section{Experimental Evaluation}\label{exp}
In our experiments, we use a commercial laptop with Intel(R) Core(TM) i9-10885H CPU and 16.0 GB RAM for the simulations, and we use a Nexys 4 DDR DIGILENT board for the FPGA XOR APUF implementations. All the APUFs or XOR APUFs being attacked are 64-bit long. Some of our source codes are modified from the GitHub repository\footnote{\label{fn:ipuf_impl}https://github.com/scluconn/DA\_PUF\_Library} of the enhanced reliability-based attack~\cite{nguyen2018interpose}. In all of our simulations, the weights of all APUFs are constructed based on the normal distribution $\mathcal{N}(0,1)$.

\subsection{Validating the Linear Relation between Non-Flipping Probability $F$ and the Delay Difference $\Delta$}

\begin{figure}[t]
    \centering
    \includegraphics[width=0.5\textwidth]{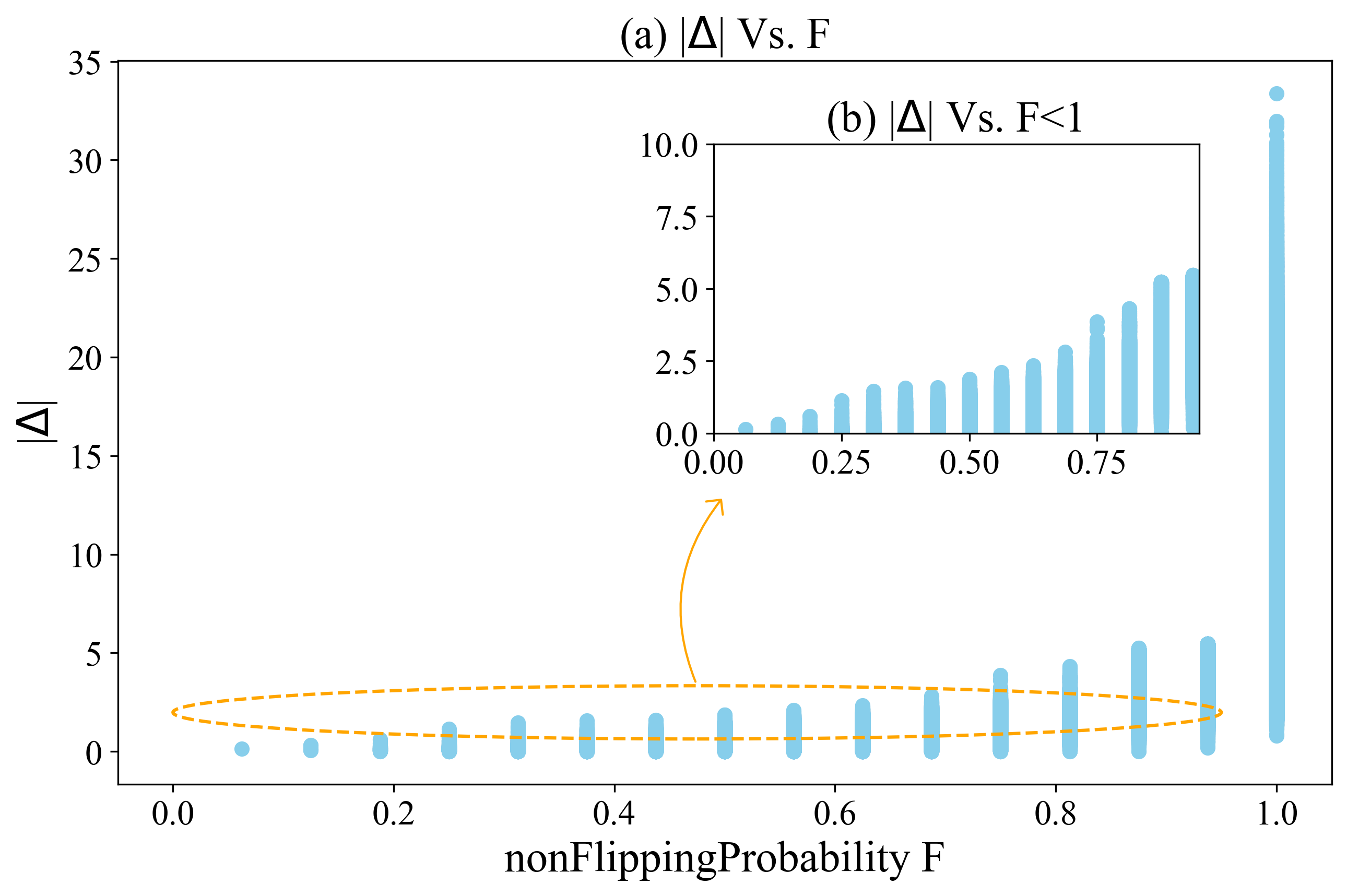}
    \caption{(a): The scatter plot of the absolute delay difference $|\Delta|$ of a random arbiter PUF versus its non-flipping probability $F$ for 300,000 CRPs. (b): Focusing on the data when the non-flipping probability $F$ is less than one.}
    \label{fig:DeltaNFPr}
\end{figure}

The main reason behind our proposed attack is the linear relation between the delay differences of an APUF and its output reliability/non-flipping probability, as we analyze in Sec.~\ref{theoritical}. %
To validate our theory (Eq.~\ref{F_linear2}) empirically, we plot the absolute value of the delay difference $|\Delta|$ and the non-flipping probability of the responses for 300,000 different centroid challenges of a random APUF. %
In Fig.~\ref{fig:DeltaNFPr}, We can see the linear growth of non-flipping probability with respect to the increasing $|\Delta|$. Fig.~\ref{fig:DeltaNFPr} (b) presents a subset of the data with $F<1$ to see the linear relationship between $|\Delta|$ and non-flipping probability more clearly. Fig.~\ref{fig:DeltaNFPr} (a) includes all the 300,000 CRPs experimental results, and because this figure includes those data with the delay difference value exceeding a certain threshold, the corresponding non-flipping probability $F$ is one; it is expected because with high $|\Delta|$ the response of an APUF becomes very reliable and not vulnerable to perturbations.

\subsection{Attacking Perfectly Reliable PUFs in Simulation}\label{exp_reli}

\begin{figure}[t]
    \centering
    \begin{tikzpicture}
	\pgfplotstableread[col sep=comma]{data.csv}\csvdata
	\pgfplotstabletranspose\datatransposed{\csvdata} 

 \begin{axis}[
    boxplot/draw direction = y,
    x axis line style = {opacity=1},
    axis x line* = bottom,
    axis y line = left,
    enlarge y limits,
    ymajorgrids,
    xtick = {1, 2, 3, 4, 5, 6, 7, 8, 9},
    xticklabel style = {align=center, font=\small, rotate=60},
    xticklabels = {$m=4$, $m=8$, $m=16$, $m=24$, $m=32$, $m=40$, $m=48$, $m=56$, $m=64$}, %
    xtick style = {draw=none}, %
    ylabel = {Modeling Accuracy},
    ytick = {0.90, 0.91, 0.92, 0.93, 0.94, 0.95, 0.96, 0.97, 0.98, 0.99}, %
]
\foreach \n in {1,...,9} {
    \addplot+[boxplot, fill, draw=black] table[y index=\n] {\datatransposed}; %
}
\end{axis}

\end{tikzpicture}
    \caption{Prediction Accuracy of the Chosen Challenge Attack on 64-bit 2-XOR APUF with different number of samples $m$.}
    \label{fig:boxplot1}

\end{figure}

\begin{table*}[]
\centering
\caption{The modeling accuracy of the individual APUFs in a simulated reliable XOR APUF.}\label{Reliable}
\begin{tabular}{|c|c|c|c|c|c|}
\hline
\#XORs&Training&Modeling Accuracy&Modeling Accuracy&{Duration}\textsuperscript{\textdagger}&{Duration}\textsuperscript{\textdaggerdbl}\\
                    &Dataset size&of our Attack&of Becker Attack~\cite{becker2015gap} $^*$& of our Attack& of Becker Attack\\ \hline
\multirow{2}{*}{1}  & \multirow{2}{*}{$20 \times 10^3$}  & \multirow{2}{*}{98.6\% - 99.0\%} & \multirow{2}{*}{98.3\% - 99.3\%} &\multirow{2}{*}{0.5 min}&\multirow{2}{*}{6 min}\\
                    &                                    &                                  &                           & &        \\ \hline
\multirow{2}{*}{4}  & \multirow{2}{*}{$150 \times 10^3$} & \multirow{2}{*}{97.6\% - 98.6\%} & \multirow{2}{*}{99.1\% - 99.7\%} &\multirow{2}{*}{4.5 min}&\multirow{2}{*}{27 min} \\
                    &                                    &                                  &                          & &         \\ \hline
\multirow{2}{*}{8}  & \multirow{2}{*}{$300 \times 10^3$} & \multirow{2}{*}{95.8\% - 98.0\%} & \multirow{2}{*}{99.1\% - 99.7\%} &\multirow{2}{*}{20 min}&\multirow{2}{*}{58 min} \\
                    &                                    &                                  &                           &&        \\ \hline
\multirow{2}{*}{16} & \multirow{2}{*}{$500 \times 10^3$} & \multirow{2}{*}{93.6\% - 94.8\%} & \multirow{2}{*}{98.7\% - 99.6\%} &\multirow{2}{*}{21 min}&\multirow{2}{*}{94 min}\\
                    &                                    &                                  &                            & &       \\ \hline
\end{tabular}
\\
\vspace{1mm}
\footnotesize{$^*$ Since Becker's attack does not work on a reliable PUF, we cannot make a fair comparison. The column fourth presents the results of Becker's attack on an \textit{unreliable} 128-bit PUF reported in~\cite{becker2015gap}.}\\
\footnotesize {\textsuperscript{\textdagger} Duration times correspond to the average duration time of each CMA-ES run.}\\
\footnotesize {\textsuperscript{\textdaggerdbl} Duration times correspond to the average time of Becker attack per run taken from~\cite{becker2015gap}. This column is not comparable with our attack duration because they were measured under different devices and different XOR APUF stages.}\\
\end{table*}

\begin{table}[t]
\centering
\caption{The modeling accuracy of our proposed attack for the individual APUFs in a simulated reliable XOR APUF.}\label{ReliableLinearScale}
\begin{tabular}{|c|c|c|c|}
\hline
\#XORs&Training&Modeling Accuracy& {Duration}\textsuperscript{\textdagger}\\
                    &Dataset size&of our Attack&\\ \hline
\multirow{2}{*}{1}  & \multirow{2}{*}{$20 \times 10^3$}  & \multirow{2}{*}{99.2\% - 99.6\%} & \multirow{2}{*}{0.5 min} \\
                    &                                    &                                  &                                  \\ \hline
\multirow{2}{*}{4}  & \multirow{2}{*}{$80 \times 10^3$} & \multirow{2}{*}{96\% - 96.5\%} & \multirow{2}{*}{3 min} \\
                    &                                    &                                  &                                  \\ \hline
\multirow{2}{*}{8}  & \multirow{2}{*}{$160 \times 10^3$} & \multirow{2}{*}{97\% - 98.5\%} & \multirow{2}{*}{6 min} \\
                    &                                    &                                  &                                  \\ \hline
\multirow{2}{*}{16} & \multirow{2}{*}{$320 \times 10^3$} & \multirow{2}{*}{94\% - 94.3\%} & \multirow{2}{*}{13 min} \\
                    &                                    &                                  &                                  \\ \hline
\end{tabular}

\footnotesize {\textsuperscript{\textdagger} Duration times correspond to the average duration time of each CMA-ES run.}\\
\end{table}

We want to study the sensitivity of the modeling accuracy w.r.t. the number of samples ($m$) around every centroid challenge. %
We test various numbers of samples on a perfectly reliable 2-XOR APUF using the proposed attack, each with $150\times 10^3$ $(\Psi, F)$ pairs. %
As shown in Fig.~\ref{fig:boxplot1}, we test the numbers of samples from 4 to 64, and we see the modeling accuracy increasing with the increasing number of samples. As the green box illustrates in Fig.~\ref{fig:boxplot1}, using 16 samples or more, the prediction accuracy of individual APUFs would be considerably high, and by increasing the number of samples from 16 to 64, there is not much remarkable improvement in accuracy. Therefore, we use 16 samples for the rest of our experiments. It seems to be a good trade-off between the modeling accuracy and the sampling/computational cost.

\vspace{1mm}
\noindent\textbf{Comparison with the reliability-based attacks}. The results in Table~\ref{Reliable} show that the proposed chosen challenge attack can attack individual APUFs in a perfectly reliable XOR APUF with high accuracy. In contrast, the reliability-based attacks do not work in the same condition. %
The prediction accuracy of the original Becker's attack (directly copied from~\cite{becker2015gap}) on noisy XOR APUFs with the same number of XORs and the same number of challenge-reliability pairs is also shown in Table~\ref{Reliable}.

\vspace{1mm}
\noindent \textbf{Scalability.} We show the scalability of our proposed attack in Table.~\ref{ReliableLinearScale}. When the size of the training dataset is linearly scaled with respect to the number of XORs, we see a slight performance degradation in the modeling accuracy, likely due to the more complex structure in larger XOR APUFs. However, the attacks on the 16-XOR APUF can still be considered successful in a practical setting.

\begin{figure}[t]
    \centering
    \includegraphics[width=0.5\textwidth]{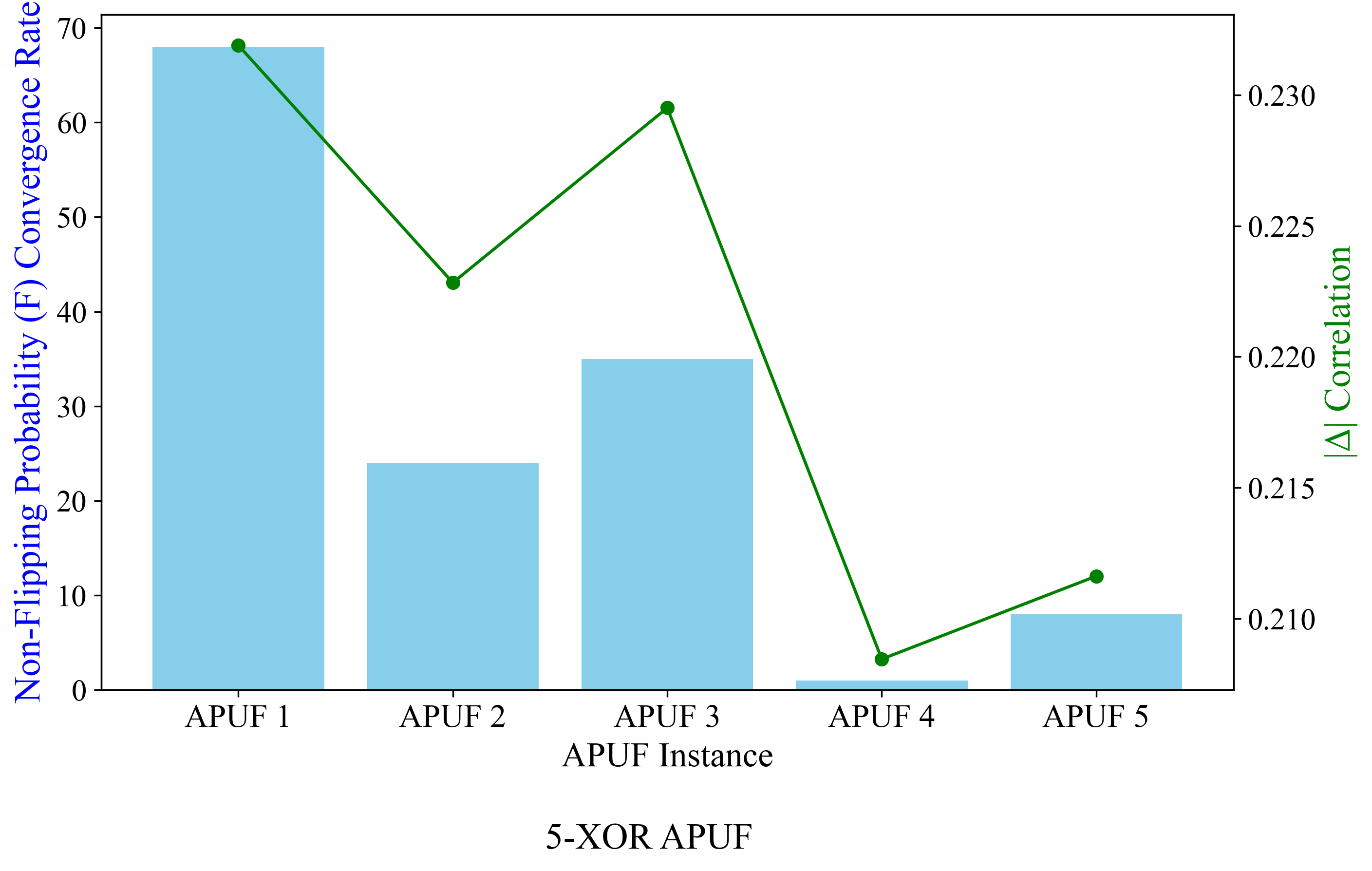}
    \caption{The right Y-axis: The correlation between the delay difference $|\Delta|$ value of constituent APUFs and the overall non-flipping probability of the 5-XOR APUF. The left Y-axis: The convergence rate of each constituent APUF. The chosen challenge attack is performed on a 5-XOR APUF with 150 CMA-ES runs using the same dataset.} 
    \label{fig:CorrDelta}
\end{figure}

\vspace{1mm}
\noindent\textbf{Convergence Rate.} Next, we study the convergence rate of each APUF in the random process of CMA-ES in our attack. The convergence rate represents the chance of an individual APUF being successfully modeled in an XOR APUF in our attack. It indicates the susceptibility of individual APUFs to the modeling attack. %
We attack a reliable 64-bit 5-XOR APUF by our attack with 150 independent CMA-ES runs using the same dataset. We can clearly see that the correlation between $|\Delta|$ of each APUF and the overall output non-flipping probability, or in short $|\Delta|$ correlation, is strongly related to the convergence rate of each APUF in Fig.~\ref{fig:CorrDelta}. The convergence rate for 150 independent CMA-ES runs is defined as the fraction of runs that successfully identify an accurate APUF model. Success is determined by achieving a Pearson correlation coefficient ($\rho$) greater than 0.95 between the predicted and observed non-flipping probabilities of the individual APUF.
When an APUF has a strong $|\Delta|$ correlation, the CMA-ES optimization process has a higher chance of convergence to that APUF model. Also, we notice that even if some of the APUFs have a relatively low $|\Delta|$ correlation in a given dataset, it is still possible for the attack to converge to them. %

\subsection{Attacking Realistically Unreliable PUFs in Simulation}\label{exp_unreli}

\begin{table*}[]
\centering
\caption{The modeling accuracy of the individual APUFs in a simulated noisy XOR APUF.  \label{unreliable}}
 \resizebox{\columnwidth}{!}{
\begin{tabular}{|c|c|c|c|c|c|c|c|}
\hline
\#XORs & Training Dataset Size& $\sigma_{Noise}$ & Modeling Accuracy& Modeling Accuracy & PUF Reliability & {Duration\textsuperscript{\textdagger}}& {Duration\textsuperscript{\textdagger}} \\ 
 & & &of our Attack& of Attack in~\cite{nguyen2018interpose}& & of our Attack&of Attack in~\cite{nguyen2018interpose} \\ \hline
\multirow{3}{*}{1}  & \multirow{2}{*}{$20 \times 10^3$}  & 0.0055           & 98.6\% - 99.6\% & 99.2\% - 99.6\%     & 99\%    & 0.6 min&0.5 min \\ \cline{3-8} 
                    &                                    & 0.01             & 99.0\% - 99.4\% & 100\% - 100\% & 98\%    & 0.6 min &0.5 min\\ \cline{3-8}
                    &                                    & 0.025            & 98.8\% - 99.5\% & 99.3\% -99.7\% & 97\%   & 1.2 min &0.5 min \\ \hline
\multirow{3}{*}{3}  & \multirow{2}{*}{$150 \times 10^3$} & 0.0055           & 97.0\% - 98.2\% & 99.2\% - 99.6\%     & 98\%  & 4.6 min &4 min  \\ \cline{3-8} 
                    &                                    & 0.01             & 98.0\% - 99.0\% & 99.6\% - 100\%      & 96\%    & 4.8 min&4 min \\ \cline{3-8}
                    &                                    & 0.025            & 98.9\% - 99.4\% & 98.8\% - 99.5\% & 92\%    & 4.8 min &4.5 min\\ \hline
\multirow{3}{*}{5}  & \multirow{2}{*}{$300 \times 10^3$} & 0.0055           & 98.8\% - 99.4\% & 98.2\% - 98.6\%     & 96\%   & 9.2 min&8.7 min  \\ \cline{3-8} 
                    &                                    & 0.01             & 97.8\% - 98.2\% & 99.4\% - 99.6\%     & 94\%   & 9.4 min&9 min  \\ \cline{3-8}
                    &                                    & 0.025            & 96.5\% - 98.8\% & 99.1\% - 99.6\% &  86\%   & 17.6 min &7.5 min \\ \hline
\multirow{3}{*}{7}  & \multirow{2}{*}{$300 \times 10^3$} & 0.0055           & 96.0\% - 98.4\% & 99.4\% - 99.8\%     & 95\%   & 10 min&9.3 min  \\ \cline{3-8} 
                    &                                    & 0.01             & 98.0\% - 98.4\% & 99.2\% - 99.4\%     & 92\%   & 11 min&9.5 min  \\ \cline{3-8}
                    &                                    & 0.025            & 96.5\% - 97.1\% & 99.6\% - 99.3\% & 82\% & 21.8 min&9.5 min \\ \hline
\multirow{3}{*}{15} & \multirow{2}{*}{$500 \times 10^3$} & 0.0055           & 93.0\% - 95.2\% & 99.4\% - 99.8\%     & 91\%  & 22 min &20 min  \\ \cline{3-8} 
                    &                                    & 0.01             & 94.2\% - 97.4\% & 99.8\% - 99.8\%     & 85\%  & 23 min &18 min  \\ \cline{3-8}
                    &                                    & 0.025            & 93.0\% - 93.8\% & 99.1\% - 99.7\% & 66\%  & 21 min  &18.5 min\\ \hline
\end{tabular}
}
\\
\vspace{1mm}
\footnotesize{$^*$ $\sigma_{noise}$ values are taken from the noise levels used in Footnote \ref{fn:ipuf_impl}}\\
\footnotesize {\textsuperscript{\textdagger} Duration times correspond to the average duration time of each CMA-ES run.}\\
\end{table*}

The proposed attack works not only with reliable PUFs but also on realistically unreliable PUFs. In every noisy APUF simulation, noisy APUF weights are independently drawn from a normal distribution with $\sigma_{Noise} * \sigma$ as the standard deviation, where $\sigma = 1$. In this way, the proposed physical error is converted to the mathematical model of APUFs in influencing the delay difference $\Delta$ of APUF in the simulation. As shown in Table~\ref{unreliable}, when $\sigma_{Noise}=0.0055$ and even with more noise as $\sigma_{Noise}=0.01$ and $\sigma_{Noise}=0.025$, the accuracy of the chosen challenge attack is considerably high, and the XOR APUF could be modeled. Note that the noise-free weight vectors of APUFs are drawn from $\mathcal{N}(0,1)$ in the simulation. The three noise level parameters $\sigma_{Noise}$ are inherited from the noisy PUF simulation\footref{fn:ipuf_impl}, and they correspond to three reliability levels measured by~\cite{nguyen2018interpose} on their APUF FPGA implementations. We also reproduced the results of the enhanced reliability-based attack on the same targets using the authors' open-source codes\footref{fn:ipuf_impl} so that the two attacks can be properly compared in Table~\ref{unreliable}. In the reliability-based attack, each challenge-reliability pair is derived from 17 measurements, just like each $(\Psi, F)$ pair is derived from 17 CRPs in our chosen challenge attack. The measured reliability of each PUF is also reported in Table~\ref{unreliable}. It is measured as the percentage of CRPs that do not flip within 10 measurements. Note that we only evaluate our attacks on odd number XOR APUFs in all the following experiments to avoid the influence of the systemic bias in the responses of even number XOR APUFs~\cite{wisiol2020short}. 

From Table~\ref{unreliable}, we notice that, in general, the proposed attack has slightly lower accuracy than the enhanced reliability-based attack. This is because the perturbation in $\Delta$ in the reliability-based attack is smaller than the perturbation introduced by one flip in $\Psi$ in our attack. Smaller perturbation means a lower chance of flipping the final response under the perturbation. This means that in the dataset collected for the chosen challenge attack, it is more likely that more than one APUF response gets flipped, and this is considered noise in our attack and negatively affects the training. %

Upon detailed examination of  Table~\ref{unreliable}, it is observed that for both attack methodologies, an increase in the standard deviation of noise from 0.0055 to 0.01 slightly enhances the modeling accuracy in general. This improvement can be potentially attributed to the expansion of the $\sigma_{Noise}$ parameter within APUF weights, which broadens the range of $\Delta$'s variations. Consequently, a greater number of challenges are likely to result in flipped responses. As demonstrated in Fig.~\ref{fig:DeltaNFPr}, the increased variability in $|\Delta|$ leads to more challenges contributing to the linear relation analysis, thereby enriching the dataset with more informative data points for more effective modeling.

\subsection{More Bit Flips Attacks in Simulation}\label{exp_2flip}

\begin{table}[t]
\centering
\caption{The modeling accuracy of the individual APUFs in a simulated XOR APUF with 4 different levels of noise under the 2-bit flip attack.}\label{2flip}
\begin{tabular}{|c|c|c|c|c|}
\hline
\#XORs             & Training               & $\sigma_{Noise}$ & Modeling & {Duration}\textsuperscript{\textdagger}\\
                   & Dataset Size           &                  &  Accuracy & \\ \hline
\multirow{4}{*}{1} & \multirow{3}{*}{$20 \times 10^3$}  & 0                & 98.4\% - 99.0\% & 1 min \\ \cline{3-5} 
                   &                                    & 0.0055           & 99.4\% - 99.8\% & 1 min  \\ \cline{3-5} 
                   &                                    & 0.01             & 99.2\% - 99.8\% & 1 min  \\ \cline{3-5}
                   &                                    & 0.025            & 98.4\% - 99.2\% & 1 min  \\ \hline
\multirow{4}{*}{3} & \multirow{3}{*}{$150 \times 10^3$} & 0                & 94.4\% - 96.0\% & 7.5 min   \\ \cline{3-5} 
                   &                                    & 0.0055           & 98.2\% - 98.8\% &10 min   \\ \cline{3-5} 
                   &                                    & 0.01             & 97.0\% - 99.2\% &7 min   \\ \cline{3-5}
                   &                                    & 0.025            & 97.6\% - 99.0\% & 7 min  \\ \hline
\multirow{4}{*}{5} & \multirow{3}{*}{$300 \times 10^3$} & 0                & 91.2\% - 93.0\% &16 min    \\ \cline{3-5} 
                   &                                    & 0.0055           & 97.4\% - 97.6\% &23 min  \\ \cline{3-5} 
                   &                                    & 0.01             & 97.0\% - 98.0\%  &15 min  \\ \cline{3-5}
                   &                                    & 0.025            & 94.6\% - 98.2\%  &15 min  \\ \hline
\multirow{4}{*}{7} & \multirow{3}{*}{$300 \times 10^3$} & 0                & 92.8\% - 95.2\%  &14 min   \\ \cline{3-5} 
                   &                                    & 0.0055           & 93.8\% - 94.6\%  &15.5 min \\ \cline{3-5} 
                   &                                    & 0.01             & 92.4\% - 92.8\%  &20   min  \\ \cline{3-5}
                   &                                    & 0.025            & 91.0\% - 96.0\% &14 min   \\ \hline
\end{tabular}

\footnotesize {\textsuperscript{\textdagger} Duration times correspond to the average duration time of each CMA-ES run.}\\
\end{table}

One may propose to use a programmable access control PUF interface~\cite{jin2022programmable} to implement a more complex access control policy, which keeps a list of erased CRPs and blocks the measurement of all CRPs of $d_H=1$ with an erased CRP. Hence, In the next step, we aim to extend the methodology of our proposed attack and investigate the resulting impact of flipping more bits in $\Psi$ on the modeling accuracy. 

We choose the challenges around the proposed challenge that have two bits flip in their corresponding $\Psi$ vector. Hence, by extending to two-bit flips, there would be $\frac{64(64-1)}{2} = 2016$ available $\Psi_i$ vectors with $d_H$=2 to be selected from, instead of just 64 vectors, if we still attack 64-bit APUFs. This extension strongly reduces the chance of using an access control interface to successfully defend against our attack because it would be too costly for an interface to block accessing all 2016 available $\Psi_i$ around an erased challenge. In other words, erasing one challenge will effectively result in erasing 2016 challenges from the PUF, and thus, the challenge space will be depleted much faster.

Table~\ref{2flip} presents the modeling accuracy of individual APUFs in an XOR APUF with four different noise levels $\sigma_{Noise}$ when 16 challenges with 2-bit flipped ($d_H=2$) in $\Psi$ are sampled around each centroid challenge. %
The results show that the XOR APUFs still could be modeled with high accuracy values. However, as expected, the accuracy is lower than that of the one-flip attacks in Tables~\ref{Reliable} and~\ref{unreliable} because likely more multiple APUF response flips occur in the training dataset, which cancels out some of the individual response flips and adds more noise into the training dataset. %
According to Table~\ref{2flip}, with the growing size of XORs, the modeling accuracy decreases, and this decrease is steeper than what occurs in Tables~\ref{Reliable} and~\ref{unreliable}. The reason is that as the size of the perturbation becomes larger through higher hamming distances, the higher the number of XORs, the higher the probability of crossing the decision boundaries of multiple APUFs.

\subsection{Compatibility of non-flipping probability-based Attack with Combined gradient-based Attack}\label{Compatibility}

\begin{table*}[t]
\caption{The modeling accuracy of combined LR-reliability attack and combined LR-nonFlipping probability attack on XOR APUFs. }\label{table_combined}
\centering
\resizebox{\columnwidth}{!}{%
\begin{tabular}{|c|c|c|c|c|c|c|c| }
 \hline
 $\#$XORs&Training Dataset Size&$\sigma_{Noise}^2$ $^\dagger$&Modeling Accuracy R~\cite{tobisch2021combining}&Modeling Accuracy $F$&PUF Reliability&$\#$Trials&$\#$Epochs\\
 \hline
 \multirow{4}{*}{4}  &\multirow{4}{*}{$20\times10^3$}  &0.5 &87.0$\%$&89.3$\%$&90$\%$&\multirow{4}{*}{6}&\multirow{4}{*}{25}\\ \cline{3-6}
                     &                                 &0.25&91.8$\%$&90.3$\%$&92$\%$&                     &                  \\ \cline{3-6}
                     &                                 &0.1&93.1$\%$ &94.4$\%$&95$\%$&                     &                  \\ \cline{3-6}
                     &                                 &0  &98.7$\%$$^*$ &97.9$\%$&100$\%$&                     &                  \\ \hline
 \multirow{4}{*}{6}  & \multirow{4}{*}{$40\times10^3$}  &0.5 &84.0$\%$&84.4$\%$&85$\%$&\multirow{4}{*}{6}&\multirow{4}{*}{25}\\ \cline{3-6}
                     &                                 &0.25&84.6$\%$&87.5$\%$&89$\%$&                     &                  \\ \cline{3-6}
                     &                                 &0.1&90.9$\%$ &90.7$\%$&93$\%$&                     &                  \\ \cline{3-6}
                     &                                 &0  &50.7$\%$$^*$ &96.0$\%$&100$\%$&                     &                  \\ \hline
 \multirow{4}{*}{8}  & \multirow{4}{*}{$60\times10^3$}  &0.5 &79.0$\%$&77.1$\%$&82$\%$&\multirow{4}{*}{6}&\multirow{4}{*}{25}\\ \cline{3-6}
                     &                                 &0.25&83.8$\%$&83.8$\%$&87$\%$&                     &                  \\ \cline{3-6}
                     &                                 &0.1&87.1$\%$ &87.9$\%$&92$\%$&                     &                  \\ \cline{3-6}
                     &                                 &0  &52.7$\%$$^*$ &94.4$\%$&100$\%$&                     &                  \\ \hline
 10 & $100\times10^3$ &0.5 &74$\%$&75.0$\%$&78$\%$&12&25\\
  \hline
 \multirow{4}{*}{10} & \multirow{4}{*}{$200\times10^3$} &0.25&79$\%$ &78.5$\%$&84$\%$ &\multirow{4}{*}{12}  &\multirow{4}{*}{40}\\ \cline{3-6}
                     &                                  &0.1 &84$\%$ &83.8$\%$&89$\%$ &                     &                   \\ \cline{3-6}
                     &                                  &0.025&89$\%$ &88.5$\%$&94$\%$ &                    &                   \\ \cline{3-6}                
                     &                                  &0   &50.3$\%$$^*$ &93.0$\%$  &100$\%$&                   &                   \\ \hline

\end{tabular}
}

\footnotesize{$^\dagger$ The PUF noise defined in~\cite{tobisch2021combining} and subsequently in section~\ref{Compatibility} is added to $\Delta$; Hence, the interpretation of the value $\sigma_{Noise}$ is slightly different from our other experiments where the noise influences every weight vector of APUFs as a structural noise. However, the reliability metric is consistent among the experimental results and offers a reference for understanding the effect of changing $\sigma_{Noise}$.}

\footnotesize{$^*$ For simulating the combined LR-Reliability attack when $\sigma_{Noise}$ is zero since computing the reliability is not possible without noise, we disable the reliability term in the loss function. The training dataset sizes (numbers of CRPs) are multiplied by 17.}
\end{table*}

\begin{table*}[t]
\caption{The modeling accuracy of combined LR-reliability attack and combined LR-nonFlipping probability attack on iPUFs. }\label{table_combinediPUF}
\centering
\resizebox{\columnwidth}{!}{%
\begin{tabular}{|c|c|c|c|c|c|c|c| }
 \hline
 (x,y)&Training Dataset Size&$\sigma_{Noise}^2$ &Modeling Accuracy R~\cite{tobisch2021combining}&Modeling Accuracy $F$&PUF Reliability&$\#$Epochs&Batch Size\\
 \hline
 \multirow{4}{*}{(1,4)}  & \multirow{4}{*}{$40\times10^3$} &0.5 &87.0$\%$&86.5$\%$&90$\%$&\multirow{4}{*}{15}&\multirow{4}{*}{256}\\ \cline{3-6}
                         &                                 &0.25&90.7$\%$&89.6$\%$&92$\%$&                     &                  \\ \cline{3-6}
                         &                                 &0.1 &92.8$\%$&93.4$\%$&95$\%$&                     &                  \\ \cline{3-6}
                         &                                 &0   &98.7$\%$$^*$   &96.2$\%$&100$\%$&                     &                  \\ \hline
 \multirow{4}{*}{(1,6)}  & \multirow{4}{*}{$150\times10^3$}&0.5 &82.0$\%$&83.3$\%$&85$\%$&\multirow{4}{*}{15}&\multirow{4}{*}{256}\\ \cline{3-6}
                         &                                 &0.25&87.8$\%$&86.8$\%$&88$\%$&                     &                  \\ \cline{3-6}
                         &                                 &0.1 &90.6$\%$&90.1$\%$&93$\%$&                     &                  \\ \cline{3-6}
                         &                                 &0   &50.1$\%$$^*$   &96.6$\%$&100$\%$&                     &                  \\ \hline
 \multirow{4}{*}{(1,8)}  & \multirow{4}{*}{$200\times10^3$}&0.5 &77.0$\%$&78.8$\%$&81$\%$&\multirow{4}{*}{25}&\multirow{4}{*}{256}\\ \cline{3-6}
                         &                                 &0.25&82.2$\%$&81.9$\%$&86$\%$&                     &                  \\ \cline{3-6}
                         &                                 &0.1 &87.5$\%$&85.9$\%$&91$\%$&                     &                  \\ \cline{3-6}
                         &                                 &0   &50.1$\%$$^*$   &93.3$\%$&100$\%$&                     &                  \\ \hline
 \multirow{3}{*}{(1,10)} & \multirow{3}{*}{$500\times10^3$}&0.5 &73.0$\%$&73.2$\%$&77$\%$&\multirow{3}{*}{25}&\multirow{3}{*}{256}\\ \cline{3-6}
                         &                                 &0.25&77.5$\%$&78$\%$&82$\%$&                     &                  \\ \cline{3-6}
                         &                                 &0.1 &83.0$\%$ &82.8$\%$&88$\%$&                     &                  \\ \hline
 \multirow{2}{*}{(1,10)} & \multirow{2}{*}{$800\times10^3$}&0.025&89.5$\%$&87.1$\%$&94$\%$&\multirow{2}{*}{25}&\multirow{2}{*}{256}\\ \cline{3-6}
                         &                                 &0    &50.5$\%$$^*$ &91.3$\%$& 100$\%$&                       &                  \\ \hline                       
\end{tabular}
}

\vspace{2mm}
\footnotesize{$^*$ For simulating the combined LR-Reliability attack when $\sigma_{Noise}$ is zero since computing the reliability is not possible without noise, we disable the reliability term in the loss function. The training dataset sizes (numbers of CRPs) are multiplied by 17.}

\end{table*}

The combined multi-objective attack framework presented in~\cite{tobisch2021combining} integrates the strengths of the direct modeling by CRPs and the reliability-based approach through a gradient-based optimization framework with multiple objectives. The authors demonstrated that combining the reliability attacks, weight constraints, and Logistic Regression (LR) into a single optimization framework enhances the efficiency of modeling attacks against strong PUFs, including XOR APUF and interpose PUF (iPUF). This combined method allows the simultaneous exploitation of both responses and reliability information, leading to a more effective attack on iPUFs compared to traditional approaches that just rely on CMA-ES or gradient-based methods without this integration. 

This combined methodology inspired us to introduce non-flipping probability into the combined attack on XOR APUF. %
Moreover, we want to make sure that our chosen challenge methodology to exploit non-flipping probability information can integrate with other objectives effectively in the combined attack framework. Furthermore, tailoring the combined attack to employ $F$ information specifically for chosen challenge attacks enables a more efficient attack when the reliability information is less accessible for the attackers. %
Our experimental investigations reveal that incorporating non-flipping probability $F$ within the combined attack methodology yields a similar level of efficiency with~\cite{tobisch2021combining} in predicting XOR APUFs behaviors across different configurations and sizes. To be precise, we use the objective of optimizing the correlation between $|\Delta|$ of individual APUFs and $F$ to replace the objective of optimizing the correlation between the measured reliability and the predicted reliability in~\cite{tobisch2021combining}.

Table~\ref{table_combined} shows the simulation results of combined LR reliability-based attack~\cite{tobisch2021combining} on XOR APUF and our proposed combined LR non-flipping probability-based attack on XOR APUFs in the same configurations and with multiple noise levels.\footnote{Our implementation of the combined attack is modified from the source codes provided by the authors of~\cite{tobisch2021combining} at https://github.com/jtobi/puf-simulation/.} This result shows the inherent stability and wide applicability of the non-flipping probability-based approach. %
 For computing non-flipping probability, 16 samples around a centroid challenge are collected. According to the same configuration, in reliability-based attacks, each challenge is measured 16+1=17 times to assess the response's reliability.
The same efficiency of the combined LR non-flipping probability-based attack with the combined LR reliability-based attack highlights the potential of the proposed method as a new standard for assessing APUF vulnerabilities.

Similar to our previous experiment, we demonstrate the feasibility of a successful attack on a reliable PUF without noise using our combined attack. As we can see in Table~\ref{table_combined}, an accuracy of 93$\%$ is produced by a combined LR non-flipping probability-based attack on 10-XOR APUF with $\sigma_{Noise}=0$. To make a fair comparison with the modeling accuracy of the combined LR-Reliability attack, we should also use the combined LR-Reliability attack to attack perfectly reliable XOR APUFs. Since it is impossible to measure reliability when $\sigma_{Noise}$=0, we disable the reliability term in the loss function of the attack, and only the response prediction accuracy is considered for this part of the experiments. Moreover, because we do not need to repeatedly measure the same CRPs on a reliable PUF, we multiply the corresponding training dataset size by the number of samples $m+1 = 17$ for a fair comparison. We see that when the number of XOR is 6 or larger, our attack achieves a significant improvement in the modeling accuracy compared with the combined LR-Reliability attack.

Furthermore, the combined LR-non-flipping probability attack targeting iPUF~\cite{nguyen2018interpose} provides the same successful prediction accuracy as the combined LR-Reliability attack when the iPUF is not reliable. Table~\ref{table_combinediPUF} includes the simulation results of the combined LR-non-flipping probability attack on iPUF along with the combined LR-Reliability attack on iPUF for comparison. To run the LR-Reliability attack on reliable iPUF, we also disable the reliability term in the loss function, and the training dataset size for these experiments is 17 times that of the corresponding LR-non-flipping probability attack. The combined LR-Reliability attack is not successful on (1,6)-iPUF and larger iPUFs when there is no noise.

In the experimental results in Tables~\ref{table_combined} and ~\ref{table_combinediPUF}, a different trend from the experimental results of previous subsections emerges concerning the relationship between the increasing noise level and the 
modeling accuracy. 
Previously, in Tables~\ref{unreliable} and~\ref{2flip}, the decrease in the noise level results in a reduction of modeling accuracy for both non-flipping probability-based and reliability-based CMA-ES ML attacks. This can be attributed to the finer sensitivity of these attacks to perturbations, which are less prevalent in a lower-noise PUF, thereby reducing the effectiveness of the modeling. Conversely, Tables ~\ref{table_combined} and ~\ref{table_combinediPUF} show that under LR-combined attack strategies, the modeling accuracy becomes higher when there is a reduction in the noise standard deviation. The reason is the direct presence of CRPs in the loss function of the combined attack. %
Thus, the type of attack and its underlying mechanics significantly influence how noise impacts modeling accuracy. The standalone non-flipping probability-based or reliability-based ML attacks suffer from reduced noise, while combined approaches benefit from it.

\subsection{Attacking FPGA Implemented PUFs}\label{exp_imp}
Last but not least, we evaluate the proposed attack on XOR APUFs implemented on an FPGA. Table~\ref{table_imp} shows the modeling accuracy of the individual APUFs under the proposed attack, and its high accuracy validates the effectiveness of our attack in practice. We target different sizes of XOR APUF up to 9 XORs. High prediction accuracy values can be achieved for constituent APUFs of XOR APUF with the chosen challenge attack. Even though the modeling accuracy falls slightly when increasing the size, it stands above 96$\%$ for the 9-XOR APUFs.

\begin{table}[t]
\caption{The modeling accuracy of the individual APUFs in a real XOR APUF on Xilinx FPGA.}\label{table_imp}
\centering
 \resizebox{0.6\columnwidth}{!}{

\begin{tabular}{|c|c|c|c|c|}
 \hline
 $\#$XORs&Training &Modeling& {CRP}&{Simulation}\\
 &Dataset Size&Accuracy&{Collection} &{Time}\textsuperscript{\textdagger}\\
 & & &{Time} &\\
 \hline
 1   & $20\times10^3$ &99.0$\%$-99.3$\%$&$\simeq 15 h$&$\simeq 0.5-1  \,$min \\
 \hline
 3   & $75\times10^3$ &98.0$\%$-98.4$\%$&$\simeq 57 h$&$\simeq 2-3  \,$ min \\
  \hline
 5& $75\times10^3$ &97.3$\%$-98.0$\%$&$\simeq 57 h$&$\simeq 2-3  \,$ min \\
  \hline
    7& $75\times10^3$ &96.6$\%$-97.6$\%$ &$\simeq 57 h$&$\simeq 2-3  \, $ min\\
  \hline
      9& $75\times10^3$ &96.2$\%$-96.3$\%$ &$\simeq 57 h$&$\simeq 2-3 \, $ min \\
  \hline
\end{tabular}
}
\\
\footnotesize {\textsuperscript{\textdagger} Simulation time is the average duration time of each CMA-ES run.}\\
\end{table}

%% file: sections/countermeasure.tex
\section{Countermeasures}\label{Countermeasures}
One straightforward but effective countermeasure against our attack is to add a one-way function (OWF) before an XOR APUF and enforce that the one-way function and the PUF must be evaluated as a whole: $R = PUF(OWF(C))$ as suggested in~\cite{gassend2002controlled,gurevin2023secure,van2023theoretical}. %
Then, an attacker cannot choose any specific PUF challenges to be evaluated anymore, due to the one-wayness of the added one-way function~\cite{katz2007introduction}. Hence, with a specific Hamming distance value, it is impossible to measure the non-flipping probability for the chosen challenge attack. 
A similar idea has been proposed in~\cite{gassend2002controlled}, which uses a hash function to control the input of the PUF and prevent selecting the challenge required for exploiting informative parameters.

However, this countermeasure may not be ideal to implement in a lightweight application scenario, e.g., RFID tags, due to the area overhead incurred by a one-way function. According to~\cite{juels2005authenticating}, there may be 1000-10000 gates on a basic RFID tag, but only 200-2000 are allocated specifically for security. The lightweight PHOTON hash function requires 865 gate equivalence, according to~\cite{maleki2017new}, which is still a relatively large portion of the whole security overhead. The occupied area for the hash function varies in different cases, but it is still considerably large compared with the small area footprint of a PUF.

Furthermore, this countermeasure does not prevent reliability-based attacks since reliability information relies only on the repetition of the measurement for the same challenge and does not depend on any specifically chosen challenges. %
Therefore, the OWF alone cannot resist the reliability-based attack on XOR APUF, and more countermeasures should be combined with it. As a strong countermeasure against the chosen challenge attack, reliability-based attacks, and combined attacks, a novel, secure, strong PUF design is needed.

\textcolor{black}{Another potential countermeasure for countering ML attacks is response obfuscation, where the PUF’s output is intentionally poisoned under hidden conditions to mislead modeling attacks. For example, Wang et al.~\cite{wang2021modeling} proposed an adversarial PUF that uses a secret trigger signal, derived from the challenge and a secret control word, to inject misleading responses. However, keeping this control word secret is not trivial and introduces overhead. Also, it should not contradict the fundamental goal of using PUF to avoid on-chip digital secrets.}

As a strong countermeasure against the chosen challenge attack, reliability-based attacks, and combined attacks, a novel, secure, strong PUF design is needed.

%% file: sections/relatedwork.tex
\section{Other Related Work}\label{relatedwork}

\noindent\textbf{Existing Chosen Challenge Attacks.}
Prior to this work, some other studies on (adaptively) chosen challenge attacks on APUF have been proposed. For example, active learning methods can improve the efficiency of modeling by collecting informative CRPs, i.e., the CRPs with the highest uncertainty~\cite{wen2018puf}. %
Adaptively choosing CRPs to measure based on optimization theory for reducing the needed number of CRPs is used in~\cite{liu2016optimization}. However, both the chosen challenge attacks above can only attack single APUFs, not XOR APUFs.

The most recent and relevant study is a chosen challenge attack exploiting the output transitions of an XOR APUF~\cite{lin2023learning}. The attack intentionally collects pairs of CRPs, which flip the response bit of an XOR APUF with limited input flips in raw challenges (instead of the $\Psi$ vectors in our attack). However, the authors directly used the collected CRPs for machine learning training to attack an XOR APUF as a whole. Thus, they only achieved up to 50\% reduction in the required number of CRPs in attacking up to 5-XOR APUFs, while our attack can attack individual APUFs instead of the XOR APUF as a whole and demonstrate successful attacks on larger XOR APUFs. 

\noindent\textbf{The Effect of Single-bit Flips in Challenges.}
The relationship of input bit flips and output bit flips of an XOR APUF has been well studied in the form of strict avalanche criteria, predictability test, and input sensitivity ~\cite{ganji2020pitfalls,lin2023learning,nguyen2016security}. However, the understanding has been mainly used for assessing the machine learning resilience of a PUF~\cite{ganji2019pufmeter,stefani2024strong} and for inspiring new PUF design ideas~\cite{ganji2021rock}. A relevant recent work analyzed the sensitivity of strong PUF responses (including XOR Arbiter PUFs, XOR Bistable Ring PUFs, and FeedForward Arbiter PUFs) to a single-bit flip in challenges as security metrics to enhance the design assessments~\cite{stefani2024strong}. In our attack, we introduce a single-bit flip in $\Psi$ instead of in the challenge and study the effect of bit flips in $\Psi$.  %

\noindent\textbf{Classical ML Attacks on XOR APUFs.} Several recent attempts have been made to attack XOR APUFs using %
classical machine learning attacks, meaning the CRPs are used directly in training~\cite{ruhrmair2010modeling,ruhrmair2013puf, wisiol2022neural ,shi2019approximation,santikellur2020computationally,santikellur2019deep}. %
A deep feedforward neural network-based modeling
attack is proposed in~\cite{santikellur2019deep} targeting Arbiter PUF and multiple compositions of APUF, including XOR-APUF up to 6 APUFs.
Wisiol et al. presented an efficient neural-network-based modeling attack on XOR APUF using challenge-response pairs~\cite{wisiol2022neural}. However, the attack does not scale well with the number of XORs in an XOR APUF. Another artificial neural network-based attack proposed by~\cite{shi2019approximation} characterizes the structure of XOR APUF up to 5 APUFs. Neural network-based tools could further reduce the computational complexity of ML attack on XOR APUF using CP-decomposition-based tensor regression network~\cite{santikellur2020computationally}.
\textcolor{black}{Hongming \textit{et al.}~\cite{hongming2024attacking} proposed a generic attack framework for modeling multiple XOR APUFs and other delay-based PUFs using a mixture-of-PUF-experts neural architecture, without structural knowledge. However, this attack requires a large number of CRPs and scales only up to 7-XOR in their experiments.}
The mentioned modeling attacks utilize passive learning methods that analyze the PUF behavior based on a set of CRPs that are collected randomly without a specific plan. %
In general, Neural networks-based attacks on XOR APUF, as detailed in~\cite{wisiol2022neural,santikellur2020computationally}, present a more general framework compared to other methods like reliability-based attacks and non-flipping probability-based attacks. However, despite their broad applicability, the complexity of these neural network methods cannot scale linearly with respect to the number of XORs and struggles with large XOR APUFs beyond a certain point. %

Recently, CalyPSO~\cite{mishra2024calypso}, an enhanced search optimization framework inspired by Particle Swarm Optimization (PSO), has been demonstrated to model delay-based PUFs effectively, including XOR APUF variants without relying on reliability or side-channel information. The CalyPSO method provides a successful attack on the largest XOR APUFs modeled to date, up to 20-XOR APUFs, but the attack still takes the XOR APUF as a whole and optimizes all the APUF models together. Moreover, their experiments were conducted on a supercomputer as Intel(R) Xeon(R) Gold 6226 CPU @ 2.70 GHz with 96 cores, 2 threads per core, 12 cores per socket, 256GB DRAM, and each experiment was spread across 4 physical cores through Python’s \textit{multiprocessing.Pool} while our research has been done on a commercial laptop, as mentioned earlier in Section~\ref{exp}.  

\textcolor{black}{A recent study~\cite{hassan2024machine} demonstrated that challenge-response obfuscation (CRO) can be bypassed using deep learning techniques. This method specifically targets PUFs that utilize CRO and uncovers both the original challenges and the behavior of the PUF by passively collecting CRPs. While their research emphasizes obfuscated versions rather than standard XOR APUFs, it exposes the weaknesses of CRO-based defenses when confronted with deep learning models. In contrast, our attack targets XOR APUFs directly using chosen-challenge queries, representing a complementary attack strategy on standard PUFs.}

Generally speaking, the existing classical PUF ML attacks that only use CRPs directly for training do not scale well with a large number of XORs. To the best of our knowledge, only the reliability-based attack and our attack can use a divide-and-conquer approach to simplify the optimization problem regardless of the size of XOR APUFs.

%% file: sections/conclusion.tex
\section{Conclusion}\label{conclusion}
The XOR APUF is a component of many state-of-the-art strong PUFs (e.g., iPUF~\cite{nguyen2018interpose}) and one of a few PUF designs deployed in practice (e.g., in RFID tags~\cite{becker2015gap}); Therefore, it is very important to analyze its security in different dimensions and conditions. In this work, we use the non-flipping probability of chosen challenges instead of the reliability for attacking individual APUFs in XOR APUFs. Our method successfully attacks different sizes of XOR APUF in both reliable and unreliable conditions with high prediction accuracy. %
The unified theoretical foundation of our attack and the reliability-based attack also shows that our attack is a generalization of the reliability-based attacks, and thus, our attack method is applicable in more scenarios, including perfectly reliable PUFs. The compatibility of the chosen challenge method with the combined multi-objective attack is another proof of the generality of non-flipping probability. %
We validate our attack experimentally using simulation and FPGA implementations.

%% file: sections/Appendix.tex
\section{Theoretical analysis of the flipping probability given a flipped $\Psi[i]$}\label{AppendixA} 
\color{black}
In this part, we investigate the probability of an APUF response flip when there is a sign flip in the feature vector $\Psi$. 

\vspace{3mm}

\noindent
{\bf Analysis of a Concrete Manufactured APUF:} Let us assume a fixed weight vector $w[\cdot]$ corresponding to a manufactured APUF.
According to Eq.~\ref{delta_eq}, the delay difference of an APUF is expressed as:
\begin{eqnarray*} \Delta &=& D + \Psi[j] \cdot w[j], \mbox{ where} \\
D&=&\sum_{i=0,\neq j}^{n} \Psi[i] \cdot w[i]. 
\end{eqnarray*}
When flipping $\Psi[j]$
from $-1$ to $+1$ or vice versa, the sign of $\Delta$ flips if and only if, for $w=w[j]$, $D+w$ and $D-w$  have a different sign. 
Recall that we assume a perfectly reliable PUF, i.e., without measurement noise\footnote{If we do assume some measurement noise, for example, in the case when we restrict access to the same challenge-response pair repeatedly, then we can merge its distribution into the distributions of $D$ and $w$ which are analyzed in the remainder of this appendix.} (otherwise, even if $D+w$ and $D-w$ have the same sign but with one of them close to zero,  measurement noise may induce a response bit flip with significant probability).
Since the sign of $\Delta$ directly corresponds to the response bit, we have a response bit flip with probability
$$
P_{{\tt flip}}= \mbox{Pr}[(D+w > 0 \,  \wedge \,   D-w < 0) \,   \vee  \, (D+w < 0 \,   \wedge \,  D-w > 0)],
$$
or equivalently $\mbox{Pr} [|w| > |D|]$.
Here, the probability is over the random selection of the challenges in the attack, and this corresponds to a uniform selection of each $\Psi[i]$ from $\{-1,+1\}$ with the exception of $\Psi[n]=1$.

According to the law of large numbers we have that $D$ is normal distributed with mean $w[n]$ and standard deviation $\hat{\sigma}$ with
$$ \hat{\sigma}^2 = 
\sum_{i=0,\neq j}^{n-1} w[i]^2.$$
Each of the delay stages $w[i]$ results from a manufacturing process that is assumed to be independent and follow the same normal distribution ${\cal N}(0,\sigma^2)$~\cite{ruhrmair2010modeling,solter2009cryptanalysis}.
Therefore, again by the law of large numbers, $\hat{\sigma}^2$ is a good estimate of $(n-1)\sigma^2$. We have $\hat{\sigma}^2 \approx (n-1)\sigma^2$ with estimation noise having a mean of $0$ and standard deviation $O(\sqrt{n})$. We conclude that approximately
$$ D-w[n] \sim {\cal N} (0, (n-1)\cdot \sigma^2).$$

\vspace{3mm}

\noindent
{\bf Expectation of the Flipping Probability:}
We can now compute the expectation  of $P_{{\tt flip}}$ over all possible realizations/manufacturing of $w=w[j]\sim {\cal N}(0,\sigma^2)$ and $v=w[n]\sim {\cal N}(0,\sigma^2)$:
\begin{eqnarray*}
   \mathbb{E}[P_{{\tt flip}}] 
 &=& \mathbb{E}[\mbox{Pr}[|w|>|D|]] \\
   &=& 
   2 \cdot \int_{v=-\infty}^{\infty} 
   \frac{e^{-\frac{v^2}{2\sigma^2}}}{\sqrt{2\pi\sigma^2}}
   \cdot \int_{D=0}^{\infty} \frac{e^{-\frac{(D-v)^2}{2(n-1){\sigma}^2}}}{\sqrt{2\pi (n-1)\sigma^2}} \\
   && \hspace{1cm} \cdot \int_{w=D}^{\infty} \frac{e^{-\frac{w^2}{2{\sigma}^2}}}{\sqrt{2\pi\sigma^2 }} \,\mbox{d}w\,\mbox{d}D \, \mbox{d}v\\
   &=& 2 \cdot  \int_{D=0}^{\infty} \frac{e^{-\frac{D^2}{2n{\sigma}^2}}}{\sqrt{2\pi n\sigma^2}} 
   \cdot \int_{w=D}^{\infty} \frac{e^{-\frac{w^2}{2{\sigma}^2}}}{\sqrt{2\pi\sigma^2 }} \, \mbox{d}w\,\mbox{d}D
\end{eqnarray*}
because adding two normal distributed random variables $v\sim {\cal N}(0,\sigma^2)$ and $D-v\sim {\cal N}(0,(n-1)\sigma^2)$ yields the random variable $D\sim {\cal N}(0,n\sigma^2)$ in the derivation.
Substitution of $w$ for $w=D \cdot x$ simplifies integration by scaling $w$ relative to $D$:
\begin{eqnarray*}\mathbb{E}[P_{{\tt flip}}] &=& 2 \cdot \int_{D=0}^{\infty} \frac{e^{-\frac{D^2}{2n{\sigma}^2}}}{\sqrt{2\pi n\sigma^2}} \cdot \int_{x=1}^{\infty} \frac{e^{-\frac{D^2 x^2}{2{\sigma}^2}}}{\sqrt{2\pi  {\sigma}^2}} \, D \, \mbox{d}x\, \mbox{d}D \\
&=& 2 \cdot \int_{x=1}^{\infty} \int_{D=0}^{\infty}  \frac{e^{-\frac{D^2}{2{\sigma}^2} (x^2 + \frac{1}{n})}}{2\pi \sqrt{n} \cdot {\sigma}^2} D \, \mbox{d}D \, dx \\
&=& \int_{x=1}^{\infty} \frac{1}{\pi \sqrt{n} (x^2 + \frac{1}{n})} dx \\
&=& \int_{x=1}^{\infty} \frac{\sqrt{n}}{\pi (nx^2 +1)} dx.
\end{eqnarray*}
Now, let $y = \sqrt{n}x$, giving:
\[\mathbb{E}[P_{\tt flip}] = \int_{y=\sqrt{n}}^{\infty} \frac{1}{\pi(y^2 + 1)} dy 
= \frac{1}{2}- \frac{1}{\pi} \cdot \arctan(\sqrt{n}).\]

\vspace{3mm}

The expected value
$\mathbb{E}[P_{{\tt flip}}]= \frac{1}{2}- \frac{1}{\pi} \cdot \arctan(\sqrt{n})$ relates inversely proportional to the expected number of challenges required for observing at least one response bit flip when flipping one $\Psi[j]$ in a $\Psi$ vector.
For $n$ equal to 64, 128, and 256, $\mathbb{E}[P_{{\tt flip}}]$ is approximately $0.0395$, $0.0280$ and $0.0198$. The expected flipping probability decreases as we increase $n$, and for larger $n$ our attack becomes less effective (unless we add some optimizations, e.g., flipping multiple $\Psi$ elements).

\vspace{3mm}

\noindent
{\bf Interpretation with respect to Attack Efficiency:}
In our attack, we select random challenges that lead to normally distributed $D$, as guaranteed by the law of large numbers. 
For a given randomly chosen challenge, when looking at the joint probability indicating which neighboring $\Psi$ among all neighboring $\Psi$ around a centroid $\Psi$ will lead to a response bit flip, each of the bit flips is positively correlated. This is because all the neighboring $\Psi$ have closely correlated $D$, which is likely to be small if any response bit flip occurs. %
Therefore, for smaller $D$, we expect to see more response bit flips.  This is exploited by our implemented CMA-ES attack.

\color{black}

\section{Flipping Position Sensitivity Study}\label{AppendixB} 
\begin{figure}[]
    \centering
    \includegraphics[width=0.5\textwidth]{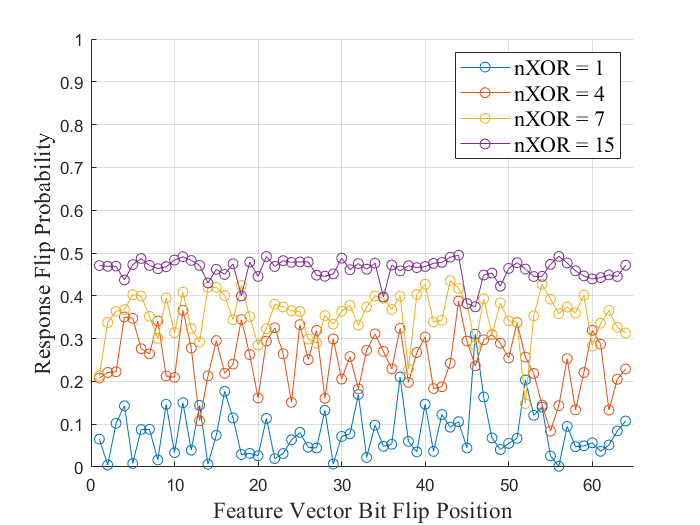}
    \caption{k-XOR APUF response flipping probability w.r.t. a single feature vector bit flip across different positions for different k-XOR configurations.}
    \label{fig:phiSensitivity}
\end{figure}

In a further experiment, we evaluated the response sensitivity to $\Psi$-bit flip across different positions. We conducted an experiment to study the response flipping probabilities based on single-bit flips in the feature vector $\Psi$. Fig.~\ref{fig:phiSensitivity} presents the response flipping probability as a function of the bit-flip positions. For each XOR configuration ($k=1, 4, 7, 15$), the response flipping probability is measured for all 64 feature vector elements ($\Psi[i]$) under the single-bit flip condition. As demonstrated, the response flipping probability remains roughly on the same level regardless of the position $i$ of the flipped $\Psi[i]$ bit. The variation of the probabilities on the different positions is due to the randomly sampled $W$ vector used in this run of the experiment. Note that this probability is different from the response flipping probability caused by single-bit flips in challenges, and they have different characteristics. For example, for a single APUF, the response flipping probability increases monotonically when the flipped challenge bit moves from the first bit to the last bit (the closest to the arbiter)~\cite{stefani2024strong}. Furthermore, as $k$ increases, the overall response flipping probability also increases, reflecting the combined effects of multiple APUFs in the XOR structure. With an increasing $k$, the variation of the overall sensitivity gets lower and its value gets closer to 50\%. This observation further demonstrates the necessity of using Pearson correlation as a robust metric against the ``noisy'' flips introduced by the other APUFs when attacking one APUF. \textcolor{black}{Theoretically, the probability of a flip in the XOR-APUF response, considering k-XOR APUFs and assuming each APUF response flip is {\bf independent}, is shown in equation~\ref{P_flip_xor}:
\begin{eqnarray}
        P_{\text{XOR\_flip}} &=& \sum_{{i = 0}, \mbox{ odd}}^{k} \binom{k}{i} \cdot P_{\text{flip}}^{i} \cdot (1 - P_{\text{flip}})^{k-i} \nonumber \\
        &=& \frac{1}{2} \cdot (1-(1-2P_{\text{flip}})^k).\label{P_flip_xor}
\end{eqnarray} 
For $k$ from 1 to 15 and $P_{\text{flip}} = 0.04$, the $P_{\text{XOR\_flip}}$ grows toward 0.5 which is validated in Fig.~\ref{fig:phiSensitivity}. }

\textcolor{black}{As a further conclusion, if we consider the case where access to the same CRP is restricted so that the classical reliability-based attack using CMA-ES does not work, we still have the measurement noise. Notably, larger $n$ leads to increased measurement noise amplified by the number of APUFs in the  XOR construction. Hence, practical XOR APUF designs are constrained to use relatively small\ignore{moderate} values of $n$ (number of APUF delay stages) and $k$ (\#XORs) to ensure stability. Accordingly, Appendix \ref{AppendixA} together with Appendix B demonstrate that our proposed attack is successful for practical implementations in access-restricted XOR APUF design.}

\section{More Comparison}\label{AppendixC}
\textcolor{black}{Table~\ref{tab:comparison} provides a comprehensive overview of the most relevant modeling attacks on XOR APUFs and related delay-based PUFs. }
\begin{table*}[t]
\scriptsize
\centering
\caption{\textcolor{black}{Comparison of Modeling Attacks on XOR APUF}}
\label{tab:comparison}
\begin{tabular}{|m{2.1cm}|>{\centering\arraybackslash}m{1.50cm}|>{\centering\arraybackslash}m{1.40cm}|>{\centering\arraybackslash}m{0.7cm}|m{2.5cm}|>{\centering\arraybackslash}m{1.3cm}|m{3.8cm}|}
\hline
\textbf{Reference} & \textbf{Training Data} & \textbf{Challenge Collection} & \textbf{Max \#XOR} & \textbf{Targeted PUFs} &\textbf{Divide and Conquer Strategy} & \textbf{Remarks} \\
\hline
Ruhrmair et al. ~\cite{ruhrmair2010modeling} & CRPs & Random & 5 &APUF, XOR APUF, FF-APUF, RO PUF &No& Evolution strategies and logistic regression on simulated CRPs\\
\hline
Ruhrmair et al. ~\cite{ruhrmair2013puf} & CRPs & Random & 5 &APUF, XOR APUF, FF-APUF, RO PUF &No& Extension of earlier work~\cite{ruhrmair2010modeling}; applied to real chips \\
\hline
Becker~\cite{becker2015gap} & Challenge-Reliability Pairs & Random & 32 & XOR APUF& Yes & Models individual APUFs in XOR APUF using Reliability-based CMA-ES \\
\hline
Liu et al. ~\cite{liu2016optimization} & CRPs & Chosen & 1 & APUF, MXbarPUF &No& Adaptive challenge selection via Chebyshev center; effective on linearly modeled PUFs \\
\hline
Wen et al. ~\cite{wen2018puf} & CRPs & Chosen & 1 & APUF&No & Active learning on APUFs; uncertainty sampling and query-by-bagging strategies \\
\hline
Santikellur et al.~\cite{santikellur2019deep} & CRPs & Random & 6 & APUF, XOR APUF, MPUFs, iPUF &No& Deep Neural Networks (NNs) on CRPs \\
\hline
Shi et al.~\cite{shi2019approximation} & CRPs & Random & 5 & XOR APUF, MPUFs&No & Logical and global approximation attack using ANNs \\
\hline
Santikellur et al.~\cite{santikellur2020computationally} & CRPs & Random & 8 & XOR APUF &No& A CP-decomposition based tensor regression NN \\
\hline
Wisiol et al. ~\cite{wisiol2022neural} & CRPs & Random & 11 & XOR APUF, FF PUF, XOR FF-APUF &No& Deep Multilayer Perceptron Attack; surpasses LR attack \\
\hline
Lin et al.~\cite{lin2023learning} & CRPs & Chosen & 5 & XOR APUF, iPUF, XOR AAPUF\textsuperscript{\textdagger}&No & Differential Chosen Challenge ML Attack for training full XOR APUF model based on output transitions.\\
\hline
Hongming et al.~\cite{hongming2024attacking} & CRPs & Random & 7 & XOR APUF, XOR FF-APUF, iPUF &No& Mixture-of-PUF-experts NN; minimal assumptions \\
\hline
Mishra et al. ~\cite{mishra2024calypso} & CRPs & Random & 20 & XOR APUF, iPUF, LP-PUF, FF-APUF, BR-PUF&No & CalyPSO: An Evolution Strategy algorithm inspired by Particle Swarm Optimization (PSO); evaluated on a supercomputing setup \\
\hline
Hassan et al. ~\cite{hassan2024machine} & CRPs & Random & 4 &$ Mn_{s1,s2,s3}$ (functionally equivalent to XOR APUF), RSO PUF\textsuperscript{\ddagger}&No &Key extraction and impersonation attack on obfuscation-based PUFs: RSO PUF and $ Mn_{s1,s2,s3}$ APUF using CMA-Es and ML \\
\hline
\textbf{Our Work} & Challenge- Non-Flipping Probability Pairs & Chosen & 15 & XOR APUF& Yes& Models individual APUFs in XOR APUF using CMA-ES; No reliability or side-channel data needed; works on reliable/unreliable PUFs \\
\hline
\end{tabular}

\footnotesize{$\ddagger$ RSO PUF: Random Set-based Obfuscation PUF,}
\footnotesize{ $\dagger$ AAPUF: Adversarial APUF}
\end{table*}